\documentclass[reprint,prl]{revtex4-1}
\usepackage{amsmath,amssymb,bm}
\usepackage{graphicx}
\usepackage{hyperref}
\usepackage{color}

\usepackage{ulem}

\begin{document}
\title{Topological Aspect of Graphene Patchwork with Regular Arrays of Nano Holes}
\date{\today}

\author{Toshikaze Kariyado}\email{Corresponding author, kariyado.toshikaze@nims.go.jp}
\author{Yongcheng Jiang}
\author{Hongxin Yang}\email{Current address: Key Laboratory of Magnetic
Materials and Devices, Ningbo Institute of Materials Technology and
Engineering, Chinese Academy of Sciences, Ningbo, 315201, China}
\author{Xiao Hu}
\email{hu.xiao@nims.go.jp}
\affiliation{International Center for Materials Nanoarchitectonics (WPI-MANA),
 National Institute for Materials Science, Tsukuba 305-0044, Japan}

  \begin{abstract}
   Triangular and honeycomb lattices are dual to each other -- if we
   puncture holes into a featureless plane in a regular triangular
   alignment, the remaining body looks like a honeycomb lattice, and
   vice versa, if the holes are in a regular honeycomb alignment, the
   remaining body has a feature of triangular lattice. In this work, we
   reveal that the electronic states in graphene sheets with nano-sized
   holes in triangular and honeycomb alignments are also dual to each
   other in a topological sense. Namely, a regular hole array perforated
   in graphene can open a band gap in the energy-momentum dispersion of
   relativistic electrons in the pristine graphene, and the insulating
   states induced by triangular and honeycomb hole arrays are distinct
   in topology. In a graphene patchwork with regions of these two hole
   arrays put side by side counterpropagating topological currents
   emerge at the domain wall. This observation indicates that the
   cerebrated atomically thin sheet is where topological physics and
   nanotechnology meet.
  \end{abstract}

 \maketitle

Electrons behave as waves in microscopic world, and a regular array of
scattering centers causes quantum interference, i.e., Bragg reflection,
which governs the electron propagation in terms of energy and momentum.
This explains band gaps and band insulators in crystals where ions are
regularly aligned. The same principle is effective even if we zoom out a
little bit: starting from Bloch waves, with which angstrom-scale
structures of the underlying crystal are already taken into account,
superstructures in micro- or meso-scopic scales can induce new band gaps
and modify the electron propagation. The first example is the
superlattice invented by Leo Esaki in order to control properties of
semiconductors \cite{esakiSL}.

In this regard, graphene -- mono atomic sheet of
carbon atoms in honeycomb structure~\cite{Novoselov:2005fk} -- is a
promising playground. First, the honeycomb array of scattering centers
is responsible for the most striking feature of graphene, emergent
relativistic fermion \cite{Geim:2007aa,RevModPhys.81.109} appearing as
an isolated gap closing point associated with linear dispersion (Dirac
cone) in the band structure. Secondly, graphene is amiable to nano
structuring \cite{Park:2008aa,PhysRevLett.101.126804,Guinea:2010aa}. One
idea is to introduce a regular array of holes, also known as antidot
lattice, into graphene, with the remaining body dubbed as graphene
nanomesh
\cite{PhysRevLett.71.4389,PhysRevLett.100.136804,doi:10.1021/ja105426h,Bai:2010aa,doi:10.1021/nl500948p,LIU201421,Kazemi:2015aa,doi:10.1021/acs.nanolett.5b04414,0957-4484-28-4-045304}. Depending
on the hole alignment, the band structure of superstructured
graphene can be either gapless or gapped, and in gapped cases the gap
size is tunable \cite{PhysRevLett.71.4389,PhysRevLett.100.136804,1367-2630-11-9-095020,Guinea5391,doi:10.1021/nn102442h,Baskin:2011aa,PhysRevB.84.125410,PhysRevB.85.115431,SMLL:SMLL201202988,Dvorak:2013aa,C4CP02090A}.

 \begin{figure*}
  \begin{center}
   \includegraphics{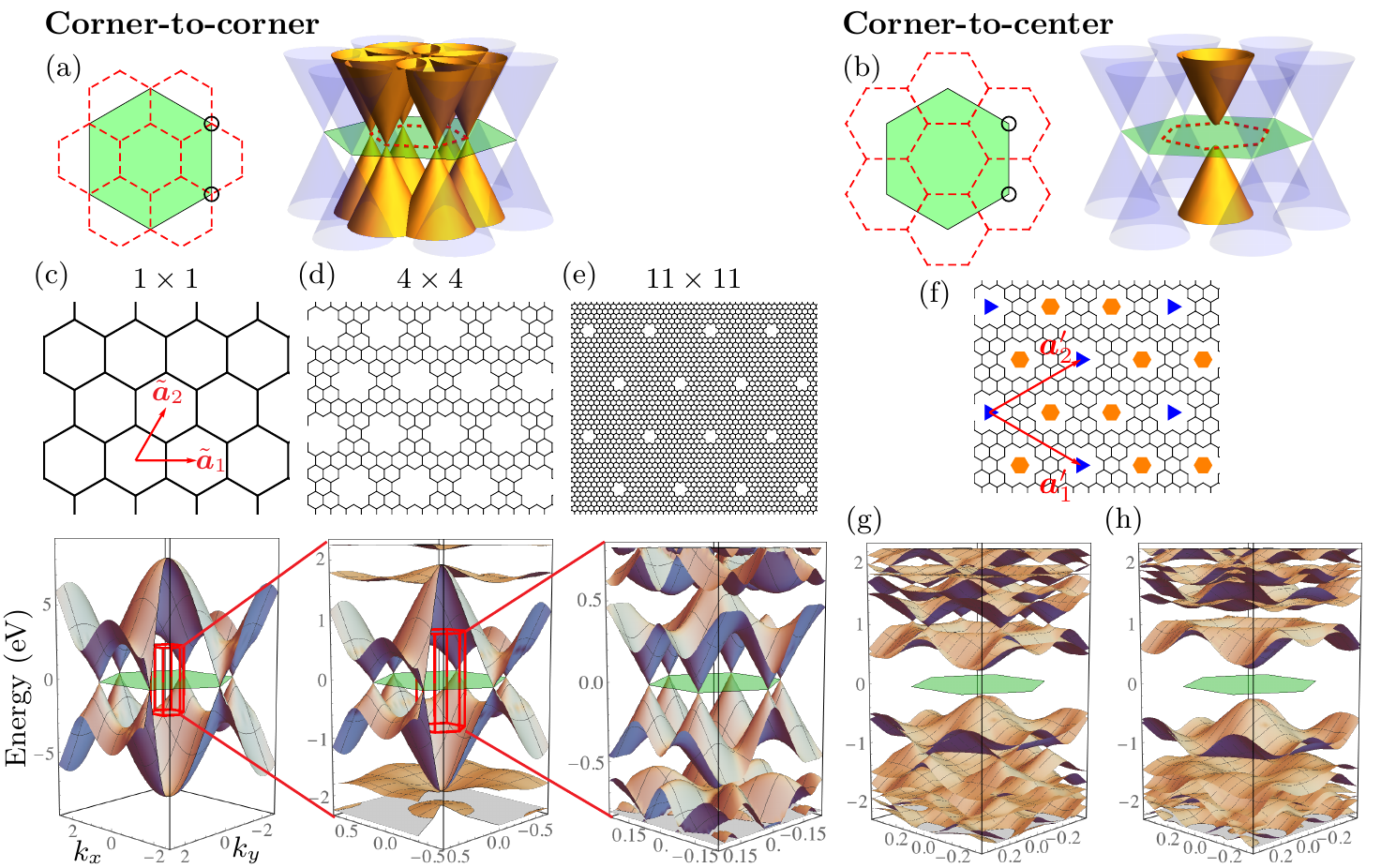}
   \caption{(a) Brillouin zone folding for $2\times 2$
   superstructure as a typical corner-to-corner folding. (b) Brillouin
   zone folding for $\sqrt{3}\times\sqrt{3}$ as a typical
   corner-to-center folding. See Ref.~\onlinecite{Guinea5391} for the
   general conditions for the corner-to-corner and corner-to-center
   foldings. (c-e) The lattice structure and the band dispersion
   with 1$\times$1 (pristine graphene), 4$\times$4, and 11$\times$11
   superstructure.  (f) Extra $\sqrt{3}\times\sqrt{3}$
   order in a 4$\times$4 superstructure. (g,h) The bulk band structure for
   two types of $4\sqrt{3}\times 4\sqrt{3}$ superstructure.}
   \label{fig:concept}
  \end{center}
 \end{figure*}
Historically, gap introduction in a honeycomb lattice model,
 or mass attachment to emergent relativistic electrons, has been
 cornerstones in discovering new topological phases of
 matter. For instance, with an appropriate time reversal symmetry (TRS)
 breaking term, the honeycomb lattice model can derive the quantum
 anomalous Hall state \cite{PhysRevLett.61.2015}, which is a typical
 topological state characterized by the Chern number
 \cite{doi:10.1080/00018732.2015.1068524}. When the
 spin-orbit coupling (SOC) is considered in a honeycomb lattice model,
 one obtains the quantum spin Hall (QSH) insulator
 \cite{PhysRevLett.95.146802,PhysRevLett.95.226801}, which is also known
 as a topological insulator specified by a $Z_2$ index. Recently, it is
 recognized that detuning the nearest-neighbor hopping integrals in the
 tight-binding model in honeycomb lattice also achieves a topological
 state characterized by \textit{mirror winding numbers}
 \cite{Wu:2016aa,Kariyado:2017aa}, i.e., a topological state
 protected by crystalline symmetry \cite{PhysRevLett.106.106802,RevModPhys.88.035005}. Yet,
 it is a challenging issue to have sizable topological gap in this
 scheme, since it involves angstrom scale manipulation in the hopping
 integrals.

 Here we propose a new strategy to realize topological states in
 graphene. Our recipe is divided into two steps, \textit{scaling} and
 \textit{gap-opening}. In the \textit{scaling} step, a superstructure
 preserving the Dirac cones at the Brillouin zone (BZ) corner is
 introduced by puncturing triangular array of nano-sized holes into
 graphene [See Fig.~\ref{fig:concept}(a)]. This process generates an
 electronic band structure similar to that of pristine graphene but with
 different length and energy scales, which permits one to work on an
 experimentally feasible length scale. In the next \textit{gap-opening}
 step, we modify the triangular hole array by filling a subset of holes
 such that the Dirac cones at the BZ corners are brought to the zone
 center [See Fig.~\ref{fig:concept}(b)] and gapped out. There are two
 kinds of superstructure in the second step, namely a triangular hole
 array and a honeycomb hole array. We reveal that graphene sheets with
 these two hole arrays are distinct in topology, and putting them side
 by side, which we call graphene patchwork, induces counterpropagating
 topological states at the domain wall. Noticing that to create
 nontrivial topology in electronic systems in terms of implanting
 nanostructures is attracting considerable current interests
 \cite{PhysRevLett.110.186601,Wang:2017aa,C7NR05325H,PhysRevLett.119.076401},
 our new finding adds a new facet in this promising derection that
 intertwines topological physics and nanotechnology.

    \medskip
    \noindent \textsf{\textbf{Model}}\\
For simplicity, we describe the electronic property of both the pristine
graphene and that with nano-sized holes in terms of a tight-binding
model on the honeycomb lattice where only the nearest neighbor hopping
$t=-2.7$ eV is taken into account \cite{PhysRevB.75.045404}. Later, this
simplification is justified by band calculations based on density
functional theory (DFT). We start with the hexagonal hole where six
carbon atoms are removed from graphene as shown in
Figs.~\ref{fig:concept}(d) and \ref{fig:concept}(e), and other
nano-sized holes will be discussed in the latter part of this paper. We
define $\tilde{\bm{a}}_{1,2}$ in Fig.~\ref{fig:concept}(c) as the unit
vectors for the pristine graphene, and the holes are arranged into a
triangular lattice with unit vectors
$\bm{a}_{1,2}=n\tilde{\bm{a}}_{1,2}$.

\medskip
\noindent \textsf{\textbf{Realizing Topologically Distinct States}}\\
It is noticed that the integer $n$ defining lattice
constant of hole array influences the
low-energy electronic structure significantly
\cite{PhysRevB.80.233405,doi:10.1021/nn200580w,doi:10.1063/1.4807426}. The
system is gapless for $n=3m\pm 1$ with integer $m$, where the band
structures of the pristine graphene and the superstructured one are
related to each other by a \textit{corner-to-corner} folding of the
Brillouin zone
\cite{Guinea5391,Dvorak:2013aa} [See Fig.~\ref{fig:concept}(a).] These
superstructures serve a good candidate for the first \textit{scaling}
step, where the low-energy band structure is similar to the band
structure of pristine graphene with certain energy-scale
renormalization, and specifically Dirac cones at the Brillouin zone (BZ)
corners are preserved, as confirmed in Figs.~\ref{fig:concept}(c), (d)
and (e), for pristine graphene, and graphene with hole arrays of
$n=3\times 1+1$ and $n=3\times 4-1$ respectively. The energy scale is
roughly proportional to the inverse of the superstructure length-scale,
reflecting the linear relation between energy and momentum of the Dirac
cone.

 In order to implement the second \textit{gap-opening} step, we
 selectively fill (i) the holes marked with
 (blue) triangles in Fig.~\ref{fig:concept}(f) leaving a honeycomb array
 of holes, or (ii) the ones with (orange) hexagons leaving a triangular
 array of holes. With
 new unit vectors $\bm{a}'_1=2\bm{a}_1-\bm{a}_2$ and
 $\bm{a}'_2=\bm{a}_1+\bm{a}_2$, this operation corresponds to a length
 scaling of $\sqrt{3}\times\sqrt{3}$ in addition to the $4\times 4$ one,
 which induces a \textit{corner-to-center} BZ folding as shown in
 Fig.~\ref{fig:concept}(b). This operation generates 
 an energy gap of 0.45 eV and 0.52 eV for the honeycomb and triangular
 hole arrays, respectively [See Figs.~\ref{fig:concept}(g) and
 \ref{fig:concept}(h)].
 
\begin{figure}[tbp]
 \begin{center}
  \includegraphics{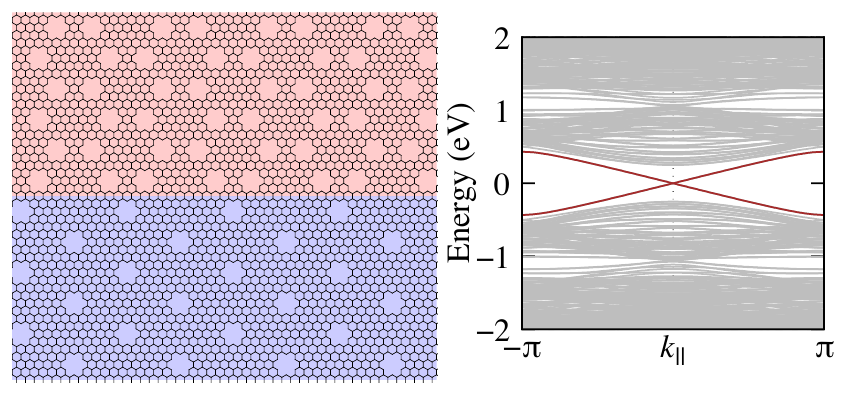}
  \caption{Left panel: Schematic picture for the graphene patchwork with
  an interface between two regions with $4\sqrt{3}\times 4\sqrt{3}$
  honeycomb and triangular hole
  arrays. Right panel: Band structure of the graphene patchwork with the
  brown dispersions representing the topological interface states.}
  \label{fig:4x4}
 \end{center}
 \end{figure}

Now we show that the two insulating states induced by the
$4\sqrt{3}\times 4\sqrt{3}$ honeycomb and triangular hole arrays are
distinct in topology by means of the interface states. We prepare a
``patchworked'' graphene sheet having two regions with the honeycomb and
 triangular hole arrays side by side as schematically illustrated in
 Fig.~\ref{fig:4x4}, and
 calculate the band structure as a function of the momentum parallel to
 the interface $k_\parallel$. 
 Figure~\ref{fig:4x4} clearly shows cross-shape interface states (brown
 lines) in the energy range of the bulk gap. This emergent dispersive
 and counterpropagating interface states resemble the helical edge state
 in QSH effect.

\medskip
\noindent \textsf{\textbf{Characterization of the Bulk Bands}}\\
The topological nature of the states under consideration
becomes clearer if we inspect the Bloch wave functions of the
\textit{bulk} states in terms of the crystalline symmetries, especially
the point group symmetries \cite{Po:2017aa,Bradlyn:2017aa,combi}. Since
the two superstructures (honeycomb and triangle) share the same $C_{6v}$
symmetry, their band topology can be specified by the numbers of states
with even parity ($N^+$) and odd parity ($N^-$) against $C_2$ rotation
(equivalent to two-dimensional spatial inversion) \cite{C2}.
 As summarized in Table~\ref{table1}, for the honeycomb hole array
 one finds in the valence bands $[N^+, N^-]= [21, 21]$ at both $\Gamma$
 and M points, specifying a topologically trivial state. In contrary,
 for the triangular array they are given by $[N^+,N^-]= [21,24]$ at
 $\Gamma$ point different from that at M point $[N^+,N^-]= [23,22]$,
 indicating a topological state. As will be discussed later, this result
 can be understood in a simple way that the order of the parity even and
 odd states at $\Gamma$ point is exchanged between the honeycomb and
 triangular hole arrays, which reverse the sign of mass attached to the
 relativistic fermion in the pristine graphene.
\begin{table}[tb]
 \centering
  \begin{tabular}{|c|c|c|}
   \hline
  structure & honeycomb & triangular \\
  \hline
   $4\sqrt{3}\times 4\sqrt{3}$& 21,21;21,21& 24,21;22,23\\
   $5\sqrt{3}\times 5\sqrt{3}$& 36,33;34,35& 36,36;36,36\\
   \hline
  \end{tabular}
 \caption{Parity indices for typical structures. The indices are
 represented as $N^+_\Gamma$,$N^-_\Gamma$;$N^+_{\text{M}}$,$N^-_{\text{M}}$,
 where $N^\pm_{\Gamma/\text{M}}$ represents the number of the valence states
 with parity $\pm$ and at $\Gamma$/M.}\label{table1}
\end{table}

  \begin{figure}[tbp]
  \begin{center}
   \includegraphics{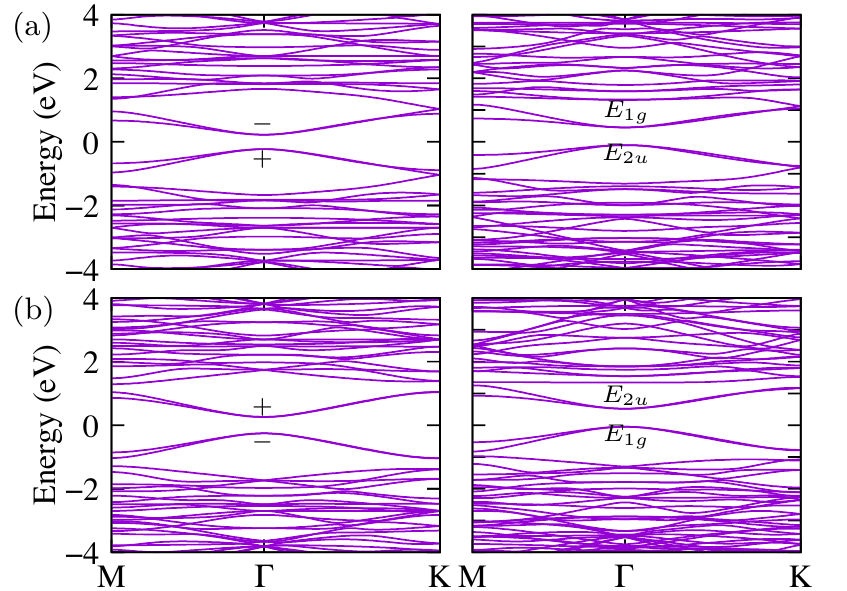}
   \caption{Bulk band structures for $4\sqrt{3}\times 4\sqrt{3}$
   superstructure of the honeycomb hole array and (b) the triangular
   hole array. The left panels derived from the tight-binding model,
   while the right panels are from DFT calculations, where the parities
   and the irreducible representations for the valence top and the
   conduction bottom are respectively denoted as well.}\label{fig:bulkband}
  \end{center}
  \end{figure}
 So far, the simple tight-binding model has been employed in our
   analysis. In order to confirm its validity, we also perform DFT
   calculations, where the perimeters of holes are terminated by
   hydrogen atoms \cite{C6RA27465J}, and band structures are evaluated
   after structural optimization. For a direct comparison, the band
   structures obtained by the tight-binding model and DFT calculations
   are displayed on the left and right column of
   Fig.~\ref{fig:bulkband}, respectively. Overall they agree well with
   each other, with the band gaps obtained by DFT calculations for both
   the honeycomb hole array (upper panels) and the triangular hole array
   (lower panels) being slightly larger than that evaluated by the
   tight-binding model. Deviations in detailed band structures are found
   away from the band gap at zero energy, which are unimportant for
   topological properties. The agreement between the two approaches is
   satisfactory considering that lattice deformations around holes
   correctly captured in DFT calculations are neglected in the simple
   tight-binding model where the hopping integral is presumed uniform in
   the whole system. Within DFT calculation, we also inspect the parity
   of wave functions. As is shown explicitly in the right column of
   Fig.~\ref{fig:bulkband}, the order of the $E_{1g}$ and $E_{2u}$
   states are exchanged between the honeycomb
   [Fig.~\ref{fig:bulkband}(a)] and triangular
   [Fig.~\ref{fig:bulkband}(b)] hole arrays, in agreement with the
   results derived with the tight-binding model as shown in the left
   column. Note that the symmetry operations in DFT calculations are 3D,
   and the states near the Fermi energy stem from
   $\pi$-orbitals. Consequently, the even/odd parity state
   $E_{1g}/E_{2u}$ corresponds to an odd/even parity state against the
   $C_2$ rotation. From these comparisons,
   we conclude that the simple tight-binding model successfully captures
   the essence of gap opening in the superstructured graphene systems,
   and the topological interface states obtained within this model as
   shown in Fig.~\ref{fig:4x4} are reliable.
   
   \begin{figure}[tbp]
    \begin{center}
     \includegraphics{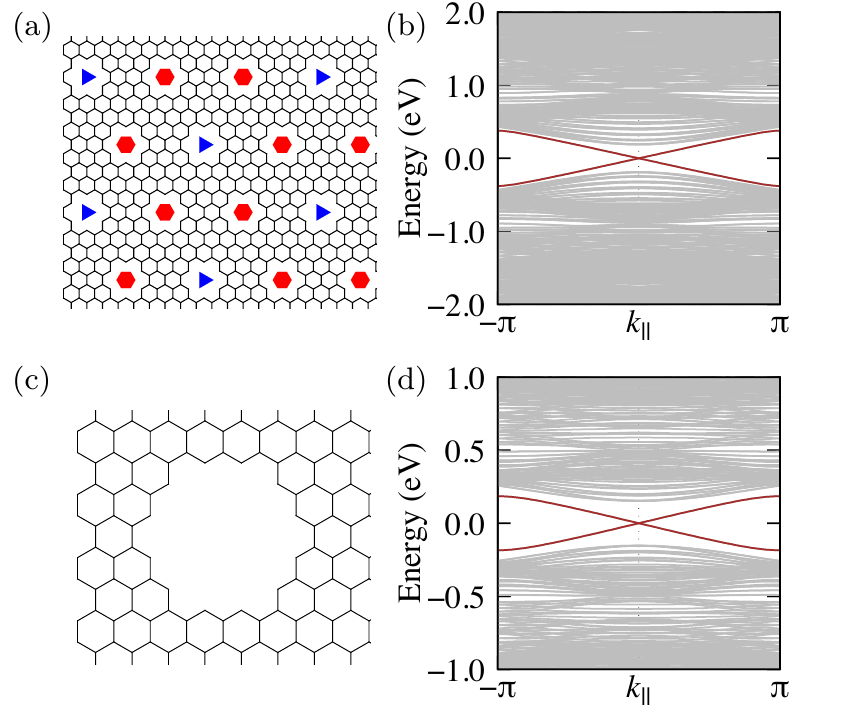}
     \caption{(a) $5\times 5$ superstructure where the triangle and
     hexagonal marks are the guides for the extra
     $\sqrt{3}\times\sqrt{3}$ superstructures. (b) The band structure for
     the interface between $5\sqrt{3}\times 5\sqrt{3}$ triangle and
     honeycomb hole arrays. (c) 24-carbon hexagonal hole. (d) The
     interface band structure for $8\sqrt{3}\times 8\sqrt{3}$ with
     24-carbon hexagonal holes.}\label{fig:variety}
    \end{center} 
   \end{figure}
   
   \medskip
   \noindent \textsf{\textbf{Discussions}}\\
   In order to see the interface states, we have used the interface along the direction of zigzag edge, for which the mirror winding
   number can be assigned with the mirror plane perpendicular to the
   interface \cite{Kariyado:2017aa}. For interfaces
   along the direction of armchair edge, a mini energy gap may open in
   the interface modes, since the mirror operation mixes the two
   sublattices of graphene \cite{Kariyado:2017aa}.
   
   One can choose length scales for generating
   graphene patchwork with topological interface states. For instance,
   let us investigate the case of $n=5$ for the \textit{scaling} process
   [Fig.~\ref{fig:variety}(a)]. In the second \textit{gap-opening} step,
   one obtains two $5\sqrt{3}\times 5\sqrt{3}$ superstructures with
   honeycomb and triangular hole arrays where band gaps are open at
   $\Gamma$ point. As can be read from Table~\ref{table1}, in valence
   bands one has for honeycomb hole array $[N^+,N^-]=[36, 33]$ at
   $\Gamma$ point, whereas $[N^+,N^-]=[34, 35]$ at M point,
   characterizing a topological state, while for triangular hole array
   $[N^+,N^-]=[36,36]$ at both $\Gamma$ and M points, specifying a
   trivial state. Because the superstructures of honeycomb and
   triangular hole arrays are distinct in topology, the graphene
   patchwork composed from them carry topological interface states as
   displayed explicitly in Fig.~\ref{fig:variety}(b). We also
   investigate other holes for superstructuring, noticing that more
   types of hole are available for choice when large unit cells are
   considered, and hole shapes may affect electronic
   states \cite{Yu2008}. As an example, we consider the case where in
   order to
   form a hole 24 carbon atoms are removed [see
   Fig.~\ref{fig:variety}(c)]. Figures~\ref{fig:variety}(d) shows the
   band structure of the corresponding graphene patchwork with
   counterpropagating interface modes. All these observations
   demonstrate the power of the \textit{scaling \& gap-opening}
   strategy.

   Triangular hole arrays of $\bm{a}_{1,2}=n\tilde{\bm{a}}_{1,2}$ with
   $n=3m$ induce a \textit{corner-to-center} BZ folding from the
   pristine graphene, and open a band gap at $\Gamma$ point. For the
   case of $n=6$, we find in valence bands $[N^+, N^-]=[15, 18]$ at
   $\Gamma$ point, whereas $[N^+, N^-]=[17, 16]$ at M
   point. However, this choice of $n$ cannot be
   used for implementing graphene patchwork, since honeycomb and
   triangular $6\sqrt{3}\times 6\sqrt{3}$ hole arrays are not
   topologically distinct, i.e., the former has $[N^+,N^-]=[51,54]$ at
   $\Gamma$ point and $[N^+,N^-]=[53,52]$ at M point, 
   while the latter has $[N^+,N^-]=[49,53]$ at $\Gamma$ point, and
   $[N^+,N^-]=[51,51]$ at M point and
   thus no interface state is
   topologically guaranteed.

   The \textit{scaling \& gap-opening} strategy allows us to design the
   topological phase in a large variety of length scale. A smaller
   superstructure yields a larger topological gap, which is advantageous
   in stability of the topological phase. However, one should keep it in
   mind that as the superstructure gets smaller, experimental
   implementations become more difficult. The small
   structures might be fabricated by bottom up methods,
   i.e., polymerizing appropriate molecules,
   but in practice, one has to figure
   out an optimal length scale to fabricate the graphene patchwork, and
   to compose a device utilizing the interface currents with topological
   protection.

     To summarize, we have demonstrated the generation of topological
    currents at the domain wall between two regions of graphene with
    different types of hole arrays, i.e., honeycomb and triangular. We have
    proposed a theoretical framework, \textit{scaling \& gap-opening}
    strategy to show the essence of the underlying physics, which can
    work as a guideline for exploration of new material phases
    by nano-structuring. While we have concentrated on perforation in
    graphene, chemical doping and passivation achieve the same goal.
    With the geometry effect playing the crucial role in our approach,
    the present idea applies for bosonic systems, such as photons and
    phonons, as well as other wave systems. The combination of
    topological physics and nanotechnology is expected to open a new era
    of fundamental and applied material science.

     \section*{Acknowledgments}
   \begin{acknowledgments}
   TK and XH thank H.~Aoki and F.~Liu for useful
    comments and discussions.
   This work was supported partially by the WPI Initiative on Materials
   Nanoarchitectonics, Ministry of Education, Cultures, Sports, Science and
   Technology, Japan, and partially by JSPS KAKENHI Grant Numbers
    JP17K14358 and JP17H02913. TK thanks the Supercomputer Center,
   the Institute for Solid State Physics, the University of Tokyo for the
   use of the facilities.
   \end{acknowledgments}

   \section*{Appendix}
  Band structures such as those shown in Fig.~2, the left column in
 Fig.~3 and Fig.~\ref{fig:variety} are obtained by diagonalizing
 numerically the hamiltonian based on the tight-binding model. For the
 right panel of Fig.~\ref{fig:4x4}, stripes of infinite length in $x$
 direction are put periodically in $y$ direction with width $16\times
 4\times (\tilde{\bm{a}}_2)_y$ for both regions of honeycomb and
 triangular hole arrays. For Fig.~\ref{fig:variety}(b) and (d), the
 width is taken as $16\times 5\times (\tilde{\bm{a}}_2)_y$ and $12\times
 8\times (\tilde{\bm{a}}_2)_y$, respectively.
 Reflecting the symmetry of the system under
 consideration, the eigen wave functions at $\Gamma$ and M points in the
 Brillouin zone can be classified by even or odd parity against the 2D
 spatial inversion, or equivalently $C_2$ rotation, with the rotation
 center at the blue (triangle) mark in Fig.~\ref{fig:concept}(f), which
 are used to specify the band topology.

  DFT calculations for band structures shown in Fig.~3 are performed
  using Vienna \textit{ab initio} simulation package
  (VASP)~\cite{PhysRevB.47.558,PhysRevB.49.14251,KRESSE199615,PhysRevB.54.11169}
  with Perdew-Becke-Erzenhof (PBE) type generalized gradient
  approximation (GGA) for exchange correlation potential
  \cite{PhysRevLett.77.3865}. The kinetic cutoff energies for the plane
  wave basis set used to expand the Kohn-Sham orbitals are 520 eV for
  the self-consistent energy calculations. 8$\times$8$\times$1 $k$-point
  meshes are used, which is sufficient to ensure good convergence in the
  total energy differences. The structural relaxations are performed
  ensuring that the Hellmann-Feynman forces acting on ions were less
  than 10$^{-3}$ eV/\AA.


\begin{thebibliography}{53}%
\makeatletter
\providecommand \@ifxundefined [1]{%
 \@ifx{#1\undefined}
}%
\providecommand \@ifnum [1]{%
 \ifnum #1\expandafter \@firstoftwo
 \else \expandafter \@secondoftwo
 \fi
}%
\providecommand \@ifx [1]{%
 \ifx #1\expandafter \@firstoftwo
 \else \expandafter \@secondoftwo
 \fi
}%
\providecommand \natexlab [1]{#1}%
\providecommand \enquote  [1]{``#1''}%
\providecommand \bibnamefont  [1]{#1}%
\providecommand \bibfnamefont [1]{#1}%
\providecommand \citenamefont [1]{#1}%
\providecommand \href@noop [0]{\@secondoftwo}%
\providecommand \href [0]{\begingroup \@sanitize@url \@href}%
\providecommand \@href[1]{\@@startlink{#1}\@@href}%
\providecommand \@@href[1]{\endgroup#1\@@endlink}%
\providecommand \@sanitize@url [0]{\catcode `\\12\catcode `\$12\catcode
  `\&12\catcode `\#12\catcode `\^12\catcode `\_12\catcode `\%12\relax}%
\providecommand \@@startlink[1]{}%
\providecommand \@@endlink[0]{}%
\providecommand \url  [0]{\begingroup\@sanitize@url \@url }%
\providecommand \@url [1]{\endgroup\@href {#1}{\urlprefix }}%
\providecommand \urlprefix  [0]{URL }%
\providecommand \Eprint [0]{\href }%
\providecommand \doibase [0]{http://dx.doi.org/}%
\providecommand \selectlanguage [0]{\@gobble}%
\providecommand \bibinfo  [0]{\@secondoftwo}%
\providecommand \bibfield  [0]{\@secondoftwo}%
\providecommand \translation [1]{[#1]}%
\providecommand \BibitemOpen [0]{}%
\providecommand \bibitemStop [0]{}%
\providecommand \bibitemNoStop [0]{.\EOS\space}%
\providecommand \EOS [0]{\spacefactor3000\relax}%
\providecommand \BibitemShut  [1]{\csname bibitem#1\endcsname}%
\let\auto@bib@innerbib\@empty
\bibitem [{\citenamefont {Esaki}\ and\ \citenamefont {Tsu}(1970)}]{esakiSL}%
  \BibitemOpen
  \bibfield  {author} {\bibinfo {author} {\bibfnamefont {L.}~\bibnamefont
  {Esaki}}\ and\ \bibinfo {author} {\bibfnamefont {R.}~\bibnamefont {Tsu}},\
  }\href@noop {} {\bibfield  {journal} {\bibinfo  {journal} {IBM J. Res. Dev.}\
  }\textbf {\bibinfo {volume} {14}},\ \bibinfo {pages} {61} (\bibinfo {year}
  {1970})}\BibitemShut {NoStop}%
\bibitem [{\citenamefont {Novoselov}\ \emph {et~al.}(2005)\citenamefont
  {Novoselov}, \citenamefont {Geim}, \citenamefont {Morozov}, \citenamefont
  {Jiang}, \citenamefont {Katsnelson}, \citenamefont {Grigorieva},
  \citenamefont {Dubonos},\ and\ \citenamefont {Firsov}}]{Novoselov:2005fk}%
  \BibitemOpen
  \bibfield  {author} {\bibinfo {author} {\bibfnamefont {K.~S.}\ \bibnamefont
  {Novoselov}}, \bibinfo {author} {\bibfnamefont {A.~K.}\ \bibnamefont {Geim}},
  \bibinfo {author} {\bibfnamefont {S.~V.}\ \bibnamefont {Morozov}}, \bibinfo
  {author} {\bibfnamefont {D.}~\bibnamefont {Jiang}}, \bibinfo {author}
  {\bibfnamefont {M.~I.}\ \bibnamefont {Katsnelson}}, \bibinfo {author}
  {\bibfnamefont {I.~V.}\ \bibnamefont {Grigorieva}}, \bibinfo {author}
  {\bibfnamefont {S.~V.}\ \bibnamefont {Dubonos}}, \ and\ \bibinfo {author}
  {\bibfnamefont {A.~A.}\ \bibnamefont {Firsov}},\ }\href@noop {} {\bibfield
  {journal} {\bibinfo  {journal} {Nature}\ }\textbf {\bibinfo {volume} {438}},\
  \bibinfo {pages} {197} (\bibinfo {year} {2005})}\BibitemShut {NoStop}%
\bibitem [{\citenamefont {Geim}\ and\ \citenamefont
  {Novoselov}(2007)}]{Geim:2007aa}%
  \BibitemOpen
  \bibfield  {author} {\bibinfo {author} {\bibfnamefont {A.~K.}\ \bibnamefont
  {Geim}}\ and\ \bibinfo {author} {\bibfnamefont {K.~S.}\ \bibnamefont
  {Novoselov}},\ }\href@noop {} {\bibfield  {journal} {\bibinfo  {journal} {Nat
  Mater}\ }\textbf {\bibinfo {volume} {6}},\ \bibinfo {pages} {183} (\bibinfo
  {year} {2007})}\BibitemShut {NoStop}%
\bibitem [{\citenamefont {Castro~Neto}\ \emph {et~al.}(2009)\citenamefont
  {Castro~Neto}, \citenamefont {Guinea}, \citenamefont {Peres}, \citenamefont
  {Novoselov},\ and\ \citenamefont {Geim}}]{RevModPhys.81.109}%
  \BibitemOpen
  \bibfield  {author} {\bibinfo {author} {\bibfnamefont {A.~H.}\ \bibnamefont
  {Castro~Neto}}, \bibinfo {author} {\bibfnamefont {F.}~\bibnamefont {Guinea}},
  \bibinfo {author} {\bibfnamefont {N.~M.~R.}\ \bibnamefont {Peres}}, \bibinfo
  {author} {\bibfnamefont {K.~S.}\ \bibnamefont {Novoselov}}, \ and\ \bibinfo
  {author} {\bibfnamefont {A.~K.}\ \bibnamefont {Geim}},\ }\href {\doibase
  10.1103/RevModPhys.81.109} {\bibfield  {journal} {\bibinfo  {journal} {Rev.
  Mod. Phys.}\ }\textbf {\bibinfo {volume} {81}},\ \bibinfo {pages} {109}
  (\bibinfo {year} {2009})}\BibitemShut {NoStop}%
\bibitem [{\citenamefont {Park}\ \emph
  {et~al.}(2008{\natexlab{a}})\citenamefont {Park}, \citenamefont {Yang},
  \citenamefont {Son}, \citenamefont {Cohen},\ and\ \citenamefont
  {Louie}}]{Park:2008aa}%
  \BibitemOpen
  \bibfield  {author} {\bibinfo {author} {\bibfnamefont {C.-H.}\ \bibnamefont
  {Park}}, \bibinfo {author} {\bibfnamefont {L.}~\bibnamefont {Yang}}, \bibinfo
  {author} {\bibfnamefont {Y.-W.}\ \bibnamefont {Son}}, \bibinfo {author}
  {\bibfnamefont {M.~L.}\ \bibnamefont {Cohen}}, \ and\ \bibinfo {author}
  {\bibfnamefont {S.~G.}\ \bibnamefont {Louie}},\ }\href
  {http://dx.doi.org/10.1038/nphys890} {\bibfield  {journal} {\bibinfo
  {journal} {Nature Physics}\ }\textbf {\bibinfo {volume} {4}},\ \bibinfo
  {pages} {213 EP } (\bibinfo {year} {2008}{\natexlab{a}})}\BibitemShut
  {NoStop}%
\bibitem [{\citenamefont {Park}\ \emph
  {et~al.}(2008{\natexlab{b}})\citenamefont {Park}, \citenamefont {Yang},
  \citenamefont {Son}, \citenamefont {Cohen},\ and\ \citenamefont
  {Louie}}]{PhysRevLett.101.126804}%
  \BibitemOpen
  \bibfield  {author} {\bibinfo {author} {\bibfnamefont {C.-H.}\ \bibnamefont
  {Park}}, \bibinfo {author} {\bibfnamefont {L.}~\bibnamefont {Yang}}, \bibinfo
  {author} {\bibfnamefont {Y.-W.}\ \bibnamefont {Son}}, \bibinfo {author}
  {\bibfnamefont {M.~L.}\ \bibnamefont {Cohen}}, \ and\ \bibinfo {author}
  {\bibfnamefont {S.~G.}\ \bibnamefont {Louie}},\ }\href {\doibase
  10.1103/PhysRevLett.101.126804} {\bibfield  {journal} {\bibinfo  {journal}
  {Phys. Rev. Lett.}\ }\textbf {\bibinfo {volume} {101}},\ \bibinfo {pages}
  {126804} (\bibinfo {year} {2008}{\natexlab{b}})}\BibitemShut {NoStop}%
\bibitem [{\citenamefont {Guinea}\ \emph {et~al.}(2010)\citenamefont {Guinea},
  \citenamefont {Katsnelson},\ and\ \citenamefont {Geim}}]{Guinea:2010aa}%
  \BibitemOpen
  \bibfield  {author} {\bibinfo {author} {\bibfnamefont {F.}~\bibnamefont
  {Guinea}}, \bibinfo {author} {\bibfnamefont {M.~I.}\ \bibnamefont
  {Katsnelson}}, \ and\ \bibinfo {author} {\bibfnamefont {A.~K.}\ \bibnamefont
  {Geim}},\ }\href@noop {} {\bibfield  {journal} {\bibinfo  {journal} {Nat.
  Phys.}\ }\textbf {\bibinfo {volume} {6}},\ \bibinfo {pages} {30} (\bibinfo
  {year} {2010})}\BibitemShut {NoStop}%
\bibitem [{\citenamefont {Shima}\ and\ \citenamefont
  {Aoki}(1993)}]{PhysRevLett.71.4389}%
  \BibitemOpen
  \bibfield  {author} {\bibinfo {author} {\bibfnamefont {N.}~\bibnamefont
  {Shima}}\ and\ \bibinfo {author} {\bibfnamefont {H.}~\bibnamefont {Aoki}},\
  }\href {\doibase 10.1103/PhysRevLett.71.4389} {\bibfield  {journal} {\bibinfo
   {journal} {Phys. Rev. Lett.}\ }\textbf {\bibinfo {volume} {71}},\ \bibinfo
  {pages} {4389} (\bibinfo {year} {1993})}\BibitemShut {NoStop}%
\bibitem [{\citenamefont {Pedersen}\ \emph {et~al.}(2008)\citenamefont
  {Pedersen}, \citenamefont {Flindt}, \citenamefont {Pedersen}, \citenamefont
  {Mortensen}, \citenamefont {Jauho},\ and\ \citenamefont
  {Pedersen}}]{PhysRevLett.100.136804}%
  \BibitemOpen
  \bibfield  {author} {\bibinfo {author} {\bibfnamefont {T.~G.}\ \bibnamefont
  {Pedersen}}, \bibinfo {author} {\bibfnamefont {C.}~\bibnamefont {Flindt}},
  \bibinfo {author} {\bibfnamefont {J.}~\bibnamefont {Pedersen}}, \bibinfo
  {author} {\bibfnamefont {N.~A.}\ \bibnamefont {Mortensen}}, \bibinfo {author}
  {\bibfnamefont {A.-P.}\ \bibnamefont {Jauho}}, \ and\ \bibinfo {author}
  {\bibfnamefont {K.}~\bibnamefont {Pedersen}},\ }\href {\doibase
  10.1103/PhysRevLett.100.136804} {\bibfield  {journal} {\bibinfo  {journal}
  {Phys. Rev. Lett.}\ }\textbf {\bibinfo {volume} {100}},\ \bibinfo {pages}
  {136804} (\bibinfo {year} {2008})}\BibitemShut {NoStop}%
\bibitem [{\citenamefont {Sinitskii}\ and\ \citenamefont
  {Tour}(2010)}]{doi:10.1021/ja105426h}%
  \BibitemOpen
  \bibfield  {author} {\bibinfo {author} {\bibfnamefont {A.}~\bibnamefont
  {Sinitskii}}\ and\ \bibinfo {author} {\bibfnamefont {J.~M.}\ \bibnamefont
  {Tour}},\ }\href {\doibase 10.1021/ja105426h} {\bibfield  {journal} {\bibinfo
   {journal} {J. Am. Chem. Soc.}\ }\textbf {\bibinfo {volume} {132}},\ \bibinfo
  {pages} {14730} (\bibinfo {year} {2010})}\BibitemShut {NoStop}%
\bibitem [{\citenamefont {Bai}\ \emph {et~al.}(2010)\citenamefont {Bai},
  \citenamefont {Zhong}, \citenamefont {Jiang}, \citenamefont {Huang},\ and\
  \citenamefont {Duan}}]{Bai:2010aa}%
  \BibitemOpen
  \bibfield  {author} {\bibinfo {author} {\bibfnamefont {J.}~\bibnamefont
  {Bai}}, \bibinfo {author} {\bibfnamefont {X.}~\bibnamefont {Zhong}}, \bibinfo
  {author} {\bibfnamefont {S.}~\bibnamefont {Jiang}}, \bibinfo {author}
  {\bibfnamefont {Y.}~\bibnamefont {Huang}}, \ and\ \bibinfo {author}
  {\bibfnamefont {X.}~\bibnamefont {Duan}},\ }\href
  {http://dx.doi.org/10.1038/nnano.2010.8} {\bibfield  {journal} {\bibinfo
  {journal} {Nat Nano}\ }\textbf {\bibinfo {volume} {5}},\ \bibinfo {pages}
  {190} (\bibinfo {year} {2010})}\BibitemShut {NoStop}%
\bibitem [{\citenamefont {Zhu}\ \emph {et~al.}(2014)\citenamefont {Zhu},
  \citenamefont {Wang}, \citenamefont {Yan}, \citenamefont {Larsen},
  \citenamefont {B{\o}ggild}, \citenamefont {Pedersen}, \citenamefont {Xiao},
  \citenamefont {Zi},\ and\ \citenamefont {Mortensen}}]{doi:10.1021/nl500948p}%
  \BibitemOpen
  \bibfield  {author} {\bibinfo {author} {\bibfnamefont {X.}~\bibnamefont
  {Zhu}}, \bibinfo {author} {\bibfnamefont {W.}~\bibnamefont {Wang}}, \bibinfo
  {author} {\bibfnamefont {W.}~\bibnamefont {Yan}}, \bibinfo {author}
  {\bibfnamefont {M.~B.}\ \bibnamefont {Larsen}}, \bibinfo {author}
  {\bibfnamefont {P.}~\bibnamefont {B{\o}ggild}}, \bibinfo {author}
  {\bibfnamefont {T.~G.}\ \bibnamefont {Pedersen}}, \bibinfo {author}
  {\bibfnamefont {S.}~\bibnamefont {Xiao}}, \bibinfo {author} {\bibfnamefont
  {J.}~\bibnamefont {Zi}}, \ and\ \bibinfo {author} {\bibfnamefont {N.~A.}\
  \bibnamefont {Mortensen}},\ }\href {\doibase 10.1021/nl500948p} {\bibfield
  {journal} {\bibinfo  {journal} {Nano Letters}\ }\textbf {\bibinfo {volume}
  {14}},\ \bibinfo {pages} {2907} (\bibinfo {year} {2014})}\BibitemShut
  {NoStop}%
\bibitem [{\citenamefont {Liu}\ \emph {et~al.}(2014)\citenamefont {Liu},
  \citenamefont {Tian}, \citenamefont {Long}, \citenamefont {Li}, \citenamefont
  {Yang}, \citenamefont {Li},\ and\ \citenamefont {Gu}}]{LIU201421}%
  \BibitemOpen
  \bibfield  {author} {\bibinfo {author} {\bibfnamefont {L.}~\bibnamefont
  {Liu}}, \bibinfo {author} {\bibfnamefont {S.}~\bibnamefont {Tian}}, \bibinfo
  {author} {\bibfnamefont {Y.}~\bibnamefont {Long}}, \bibinfo {author}
  {\bibfnamefont {W.}~\bibnamefont {Li}}, \bibinfo {author} {\bibfnamefont
  {H.}~\bibnamefont {Yang}}, \bibinfo {author} {\bibfnamefont {J.}~\bibnamefont
  {Li}}, \ and\ \bibinfo {author} {\bibfnamefont {C.}~\bibnamefont {Gu}},\
  }\href {\doibase http://dx.doi.org/10.1016/j.vacuum.2014.01.015} {\bibfield
  {journal} {\bibinfo  {journal} {Vacuum}\ }\textbf {\bibinfo {volume} {105}},\
  \bibinfo {pages} {21 } (\bibinfo {year} {2014})}\BibitemShut {NoStop}%
\bibitem [{\citenamefont {Kazemi}\ \emph {et~al.}(2015)\citenamefont {Kazemi},
  \citenamefont {He}, \citenamefont {Alaie}, \citenamefont {Ghasemi},
  \citenamefont {Dawson}, \citenamefont {Cavallo}, \citenamefont {Habteyes},
  \citenamefont {Brueck},\ and\ \citenamefont {Krishna}}]{Kazemi:2015aa}%
  \BibitemOpen
  \bibfield  {author} {\bibinfo {author} {\bibfnamefont {A.}~\bibnamefont
  {Kazemi}}, \bibinfo {author} {\bibfnamefont {X.}~\bibnamefont {He}}, \bibinfo
  {author} {\bibfnamefont {S.}~\bibnamefont {Alaie}}, \bibinfo {author}
  {\bibfnamefont {J.}~\bibnamefont {Ghasemi}}, \bibinfo {author} {\bibfnamefont
  {N.~M.}\ \bibnamefont {Dawson}}, \bibinfo {author} {\bibfnamefont
  {F.}~\bibnamefont {Cavallo}}, \bibinfo {author} {\bibfnamefont {T.~G.}\
  \bibnamefont {Habteyes}}, \bibinfo {author} {\bibfnamefont {S.~R.~J.}\
  \bibnamefont {Brueck}}, \ and\ \bibinfo {author} {\bibfnamefont
  {S.}~\bibnamefont {Krishna}},\ }\href {http://dx.doi.org/10.1038/srep11463}
  {\bibfield  {journal} {\bibinfo  {journal} {Sci. Rep.}\ }\textbf {\bibinfo
  {volume} {5}},\ \bibinfo {pages} {11463 EP } (\bibinfo {year}
  {2015})}\BibitemShut {NoStop}%
\bibitem [{\citenamefont {Sandner}\ \emph {et~al.}(2015)\citenamefont
  {Sandner}, \citenamefont {Preis}, \citenamefont {Schell}, \citenamefont
  {Giudici}, \citenamefont {Watanabe}, \citenamefont {Taniguchi}, \citenamefont
  {Weiss},\ and\ \citenamefont {Eroms}}]{doi:10.1021/acs.nanolett.5b04414}%
  \BibitemOpen
  \bibfield  {author} {\bibinfo {author} {\bibfnamefont {A.}~\bibnamefont
  {Sandner}}, \bibinfo {author} {\bibfnamefont {T.}~\bibnamefont {Preis}},
  \bibinfo {author} {\bibfnamefont {C.}~\bibnamefont {Schell}}, \bibinfo
  {author} {\bibfnamefont {P.}~\bibnamefont {Giudici}}, \bibinfo {author}
  {\bibfnamefont {K.}~\bibnamefont {Watanabe}}, \bibinfo {author}
  {\bibfnamefont {T.}~\bibnamefont {Taniguchi}}, \bibinfo {author}
  {\bibfnamefont {D.}~\bibnamefont {Weiss}}, \ and\ \bibinfo {author}
  {\bibfnamefont {J.}~\bibnamefont {Eroms}},\ }\href {\doibase
  10.1021/acs.nanolett.5b04414} {\bibfield  {journal} {\bibinfo  {journal}
  {Nano Letters}\ }\textbf {\bibinfo {volume} {15}},\ \bibinfo {pages} {8402}
  (\bibinfo {year} {2015})}\BibitemShut {NoStop}%
\bibitem [{\citenamefont {Gao}\ \emph {et~al.}(2017)\citenamefont {Gao},
  \citenamefont {Liu}, \citenamefont {Ye}, \citenamefont {Sui}, \citenamefont
  {Yan}, \citenamefont {Cai}, \citenamefont {Lv}, \citenamefont {Li},
  \citenamefont {Chen}, \citenamefont {Zheng},\ and\ \citenamefont
  {Shi}}]{0957-4484-28-4-045304}%
  \BibitemOpen
  \bibfield  {author} {\bibinfo {author} {\bibfnamefont {F.}~\bibnamefont
  {Gao}}, \bibinfo {author} {\bibfnamefont {F.}~\bibnamefont {Liu}}, \bibinfo
  {author} {\bibfnamefont {Z.}~\bibnamefont {Ye}}, \bibinfo {author}
  {\bibfnamefont {C.}~\bibnamefont {Sui}}, \bibinfo {author} {\bibfnamefont
  {B.}~\bibnamefont {Yan}}, \bibinfo {author} {\bibfnamefont {P.}~\bibnamefont
  {Cai}}, \bibinfo {author} {\bibfnamefont {B.}~\bibnamefont {Lv}}, \bibinfo
  {author} {\bibfnamefont {Y.}~\bibnamefont {Li}}, \bibinfo {author}
  {\bibfnamefont {N.}~\bibnamefont {Chen}}, \bibinfo {author} {\bibfnamefont
  {Y.}~\bibnamefont {Zheng}}, \ and\ \bibinfo {author} {\bibfnamefont
  {Y.}~\bibnamefont {Shi}},\ }\href
  {http://stacks.iop.org/0957-4484/28/i=4/a=045304} {\bibfield  {journal}
  {\bibinfo  {journal} {Nanotechnology}\ }\textbf {\bibinfo {volume} {28}},\
  \bibinfo {pages} {045304} (\bibinfo {year} {2017})}\BibitemShut {NoStop}%
\bibitem [{\citenamefont {F{\"u}rst}\ \emph {et~al.}(2009)\citenamefont
  {F{\"u}rst}, \citenamefont {Pedersen}, \citenamefont {Flindt}, \citenamefont
  {Mortensen}, \citenamefont {Brandbyge}, \citenamefont {Pedersen},\ and\
  \citenamefont {Jauho}}]{1367-2630-11-9-095020}%
  \BibitemOpen
  \bibfield  {author} {\bibinfo {author} {\bibfnamefont {J.~A.}\ \bibnamefont
  {F{\"u}rst}}, \bibinfo {author} {\bibfnamefont {J.~G.}\ \bibnamefont
  {Pedersen}}, \bibinfo {author} {\bibfnamefont {C.}~\bibnamefont {Flindt}},
  \bibinfo {author} {\bibfnamefont {N.~A.}\ \bibnamefont {Mortensen}}, \bibinfo
  {author} {\bibfnamefont {M.}~\bibnamefont {Brandbyge}}, \bibinfo {author}
  {\bibfnamefont {T.~G.}\ \bibnamefont {Pedersen}}, \ and\ \bibinfo {author}
  {\bibfnamefont {A.-P.}\ \bibnamefont {Jauho}},\ }\href
  {http://stacks.iop.org/1367-2630/11/i=9/a=095020} {\bibfield  {journal}
  {\bibinfo  {journal} {New J. Phys.}\ }\textbf {\bibinfo {volume} {11}},\
  \bibinfo {pages} {095020} (\bibinfo {year} {2009})}\BibitemShut {NoStop}%
\bibitem [{\citenamefont {Guinea}\ and\ \citenamefont
  {Low}(2010)}]{Guinea5391}%
  \BibitemOpen
  \bibfield  {author} {\bibinfo {author} {\bibfnamefont {F.}~\bibnamefont
  {Guinea}}\ and\ \bibinfo {author} {\bibfnamefont {T.}~\bibnamefont {Low}},\
  }\href {\doibase 10.1098/rsta.2010.0214} {\bibfield  {journal} {\bibinfo
  {journal} {Philosophical Transactions of the Royal Society of London A:
  Mathematical, Physical and Engineering Sciences}\ }\textbf {\bibinfo {volume}
  {368}},\ \bibinfo {pages} {5391} (\bibinfo {year} {2010})}\BibitemShut
  {NoStop}%
\bibitem [{\citenamefont {Petersen}\ \emph {et~al.}(2011)\citenamefont
  {Petersen}, \citenamefont {Pedersen},\ and\ \citenamefont
  {Jauho}}]{doi:10.1021/nn102442h}%
  \BibitemOpen
  \bibfield  {author} {\bibinfo {author} {\bibfnamefont {R.}~\bibnamefont
  {Petersen}}, \bibinfo {author} {\bibfnamefont {T.~G.}\ \bibnamefont
  {Pedersen}}, \ and\ \bibinfo {author} {\bibfnamefont {A.-P.}\ \bibnamefont
  {Jauho}},\ }\href {\doibase 10.1021/nn102442h} {\bibfield  {journal}
  {\bibinfo  {journal} {ACS Nano}\ }\textbf {\bibinfo {volume} {5}},\ \bibinfo
  {pages} {523} (\bibinfo {year} {2011})}\BibitemShut {NoStop}%
\bibitem [{\citenamefont {Baskin}\ and\ \citenamefont
  {Kr{\'a}l}(2011)}]{Baskin:2011aa}%
  \BibitemOpen
  \bibfield  {author} {\bibinfo {author} {\bibfnamefont {A.}~\bibnamefont
  {Baskin}}\ and\ \bibinfo {author} {\bibfnamefont {P.}~\bibnamefont
  {Kr{\'a}l}},\ }\href {http://dx.doi.org/10.1038/srep00036} {\bibfield
  {journal} {\bibinfo  {journal} {Sci. Rep.}\ }\textbf {\bibinfo {volume}
  {1}},\ \bibinfo {pages} {36 EP } (\bibinfo {year} {2011})}\BibitemShut
  {NoStop}%
\bibitem [{\citenamefont {Cui}\ \emph {et~al.}(2011)\citenamefont {Cui},
  \citenamefont {Zheng}, \citenamefont {Liu}, \citenamefont {Li}, \citenamefont
  {Delley}, \citenamefont {Stampfl},\ and\ \citenamefont
  {Ringer}}]{PhysRevB.84.125410}%
  \BibitemOpen
  \bibfield  {author} {\bibinfo {author} {\bibfnamefont {X.~Y.}\ \bibnamefont
  {Cui}}, \bibinfo {author} {\bibfnamefont {R.~K.}\ \bibnamefont {Zheng}},
  \bibinfo {author} {\bibfnamefont {Z.~W.}\ \bibnamefont {Liu}}, \bibinfo
  {author} {\bibfnamefont {L.}~\bibnamefont {Li}}, \bibinfo {author}
  {\bibfnamefont {B.}~\bibnamefont {Delley}}, \bibinfo {author} {\bibfnamefont
  {C.}~\bibnamefont {Stampfl}}, \ and\ \bibinfo {author} {\bibfnamefont
  {S.~P.}\ \bibnamefont {Ringer}},\ }\href {\doibase
  10.1103/PhysRevB.84.125410} {\bibfield  {journal} {\bibinfo  {journal} {Phys.
  Rev. B}\ }\textbf {\bibinfo {volume} {84}},\ \bibinfo {pages} {125410}
  (\bibinfo {year} {2011})}\BibitemShut {NoStop}%
\bibitem [{\citenamefont {Oswald}\ and\ \citenamefont
  {Wu}(2012)}]{PhysRevB.85.115431}%
  \BibitemOpen
  \bibfield  {author} {\bibinfo {author} {\bibfnamefont {W.}~\bibnamefont
  {Oswald}}\ and\ \bibinfo {author} {\bibfnamefont {Z.}~\bibnamefont {Wu}},\
  }\href {\doibase 10.1103/PhysRevB.85.115431} {\bibfield  {journal} {\bibinfo
  {journal} {Phys. Rev. B}\ }\textbf {\bibinfo {volume} {85}},\ \bibinfo
  {pages} {115431} (\bibinfo {year} {2012})}\BibitemShut {NoStop}%
\bibitem [{\citenamefont {Liu}\ \emph {et~al.}(2013)\citenamefont {Liu},
  \citenamefont {Zhang},\ and\ \citenamefont {Guo}}]{SMLL:SMLL201202988}%
  \BibitemOpen
  \bibfield  {author} {\bibinfo {author} {\bibfnamefont {X.}~\bibnamefont
  {Liu}}, \bibinfo {author} {\bibfnamefont {Z.}~\bibnamefont {Zhang}}, \ and\
  \bibinfo {author} {\bibfnamefont {W.}~\bibnamefont {Guo}},\ }\href {\doibase
  10.1002/smll.201202988} {\bibfield  {journal} {\bibinfo  {journal} {Small}\
  }\textbf {\bibinfo {volume} {9}},\ \bibinfo {pages} {1405} (\bibinfo {year}
  {2013})}\BibitemShut {NoStop}%
\bibitem [{\citenamefont {Dvorak}\ \emph {et~al.}(2013)\citenamefont {Dvorak},
  \citenamefont {Oswald},\ and\ \citenamefont {Wu}}]{Dvorak:2013aa}%
  \BibitemOpen
  \bibfield  {author} {\bibinfo {author} {\bibfnamefont {M.}~\bibnamefont
  {Dvorak}}, \bibinfo {author} {\bibfnamefont {W.}~\bibnamefont {Oswald}}, \
  and\ \bibinfo {author} {\bibfnamefont {Z.}~\bibnamefont {Wu}},\ }\href
  {http://dx.doi.org/10.1038/srep02289} {\bibfield  {journal} {\bibinfo
  {journal} {Sci. Rep.}\ }\textbf {\bibinfo {volume} {3}},\ \bibinfo {pages}
  {2289 EP } (\bibinfo {year} {2013})}\BibitemShut {NoStop}%
\bibitem [{\citenamefont {Ouyang}\ \emph {et~al.}(2014)\citenamefont {Ouyang},
  \citenamefont {Peng}, \citenamefont {Yang}, \citenamefont {Chen},
  \citenamefont {Zou},\ and\ \citenamefont {Xiong}}]{C4CP02090A}%
  \BibitemOpen
  \bibfield  {author} {\bibinfo {author} {\bibfnamefont {F.}~\bibnamefont
  {Ouyang}}, \bibinfo {author} {\bibfnamefont {S.}~\bibnamefont {Peng}},
  \bibinfo {author} {\bibfnamefont {Z.}~\bibnamefont {Yang}}, \bibinfo {author}
  {\bibfnamefont {Y.}~\bibnamefont {Chen}}, \bibinfo {author} {\bibfnamefont
  {H.}~\bibnamefont {Zou}}, \ and\ \bibinfo {author} {\bibfnamefont
  {X.}~\bibnamefont {Xiong}},\ }\href {\doibase 10.1039/C4CP02090A} {\bibfield
  {journal} {\bibinfo  {journal} {Phys. Chem. Chem. Phys.}\ }\textbf {\bibinfo
  {volume} {16}},\ \bibinfo {pages} {20524} (\bibinfo {year}
  {2014})}\BibitemShut {NoStop}%
\bibitem [{\citenamefont {Haldane}(1988)}]{PhysRevLett.61.2015}%
  \BibitemOpen
  \bibfield  {author} {\bibinfo {author} {\bibfnamefont {F.~D.~M.}\
  \bibnamefont {Haldane}},\ }\href {\doibase 10.1103/PhysRevLett.61.2015}
  {\bibfield  {journal} {\bibinfo  {journal} {Phys. Rev. Lett.}\ }\textbf
  {\bibinfo {volume} {61}},\ \bibinfo {pages} {2015} (\bibinfo {year}
  {1988})}\BibitemShut {NoStop}%
\bibitem [{\citenamefont {Weng}\ \emph {et~al.}(2015)\citenamefont {Weng},
  \citenamefont {Yu}, \citenamefont {Hu}, \citenamefont {Dai},\ and\
  \citenamefont {Fang}}]{doi:10.1080/00018732.2015.1068524}%
  \BibitemOpen
  \bibfield  {author} {\bibinfo {author} {\bibfnamefont {H.}~\bibnamefont
  {Weng}}, \bibinfo {author} {\bibfnamefont {R.}~\bibnamefont {Yu}}, \bibinfo
  {author} {\bibfnamefont {X.}~\bibnamefont {Hu}}, \bibinfo {author}
  {\bibfnamefont {X.}~\bibnamefont {Dai}}, \ and\ \bibinfo {author}
  {\bibfnamefont {Z.}~\bibnamefont {Fang}},\ }\href {\doibase
  10.1080/00018732.2015.1068524} {\bibfield  {journal} {\bibinfo  {journal}
  {Adv. Phys.}\ }\textbf {\bibinfo {volume} {64}},\ \bibinfo {pages} {227}
  (\bibinfo {year} {2015})}\BibitemShut {NoStop}%
\bibitem [{\citenamefont {Kane}\ and\ \citenamefont
  {Mele}(2005{\natexlab{a}})}]{PhysRevLett.95.146802}%
  \BibitemOpen
  \bibfield  {author} {\bibinfo {author} {\bibfnamefont {C.~L.}\ \bibnamefont
  {Kane}}\ and\ \bibinfo {author} {\bibfnamefont {E.~J.}\ \bibnamefont
  {Mele}},\ }\href {\doibase 10.1103/PhysRevLett.95.146802} {\bibfield
  {journal} {\bibinfo  {journal} {Phys. Rev. Lett.}\ }\textbf {\bibinfo
  {volume} {95}},\ \bibinfo {pages} {146802} (\bibinfo {year}
  {2005}{\natexlab{a}})}\BibitemShut {NoStop}%
\bibitem [{\citenamefont {Kane}\ and\ \citenamefont
  {Mele}(2005{\natexlab{b}})}]{PhysRevLett.95.226801}%
  \BibitemOpen
  \bibfield  {author} {\bibinfo {author} {\bibfnamefont {C.~L.}\ \bibnamefont
  {Kane}}\ and\ \bibinfo {author} {\bibfnamefont {E.~J.}\ \bibnamefont
  {Mele}},\ }\href {\doibase 10.1103/PhysRevLett.95.226801} {\bibfield
  {journal} {\bibinfo  {journal} {Phys. Rev. Lett.}\ }\textbf {\bibinfo
  {volume} {95}},\ \bibinfo {pages} {226801} (\bibinfo {year}
  {2005}{\natexlab{b}})}\BibitemShut {NoStop}%
\bibitem [{\citenamefont {Wu}\ and\ \citenamefont {Hu}(2016)}]{Wu:2016aa}%
  \BibitemOpen
  \bibfield  {author} {\bibinfo {author} {\bibfnamefont {L.-H.}\ \bibnamefont
  {Wu}}\ and\ \bibinfo {author} {\bibfnamefont {X.}~\bibnamefont {Hu}},\
  }\href@noop {} {\bibfield  {journal} {\bibinfo  {journal} {Sci. Rep.}\
  }\textbf {\bibinfo {volume} {6}},\ \bibinfo {pages} {24347} (\bibinfo {year}
  {2016})}\BibitemShut {NoStop}%
\bibitem [{\citenamefont {Kariyado}\ and\ \citenamefont
  {Hu}(2017)}]{Kariyado:2017aa}%
  \BibitemOpen
  \bibfield  {author} {\bibinfo {author} {\bibfnamefont {T.}~\bibnamefont
  {Kariyado}}\ and\ \bibinfo {author} {\bibfnamefont {X.}~\bibnamefont {Hu}},\
  }\href {\doibase 10.1038/s41598-017-16334-0} {\bibfield  {journal} {\bibinfo
  {journal} {Sci. Rep.}\ }\textbf {\bibinfo {volume} {7}},\ \bibinfo {pages}
  {16515} (\bibinfo {year} {2017})}\BibitemShut {NoStop}%
\bibitem [{\citenamefont {Fu}(2011)}]{PhysRevLett.106.106802}%
  \BibitemOpen
  \bibfield  {author} {\bibinfo {author} {\bibfnamefont {L.}~\bibnamefont
  {Fu}},\ }\href {\doibase 10.1103/PhysRevLett.106.106802} {\bibfield
  {journal} {\bibinfo  {journal} {Phys. Rev. Lett.}\ }\textbf {\bibinfo
  {volume} {106}},\ \bibinfo {pages} {106802} (\bibinfo {year}
  {2011})}\BibitemShut {NoStop}%
\bibitem [{\citenamefont {Chiu}\ \emph {et~al.}(2016)\citenamefont {Chiu},
  \citenamefont {Teo}, \citenamefont {Schnyder},\ and\ \citenamefont
  {Ryu}}]{RevModPhys.88.035005}%
  \BibitemOpen
  \bibfield  {author} {\bibinfo {author} {\bibfnamefont {C.-K.}\ \bibnamefont
  {Chiu}}, \bibinfo {author} {\bibfnamefont {J.~C.~Y.}\ \bibnamefont {Teo}},
  \bibinfo {author} {\bibfnamefont {A.~P.}\ \bibnamefont {Schnyder}}, \ and\
  \bibinfo {author} {\bibfnamefont {S.}~\bibnamefont {Ryu}},\ }\href {\doibase
  10.1103/RevModPhys.88.035005} {\bibfield  {journal} {\bibinfo  {journal}
  {Rev. Mod. Phys.}\ }\textbf {\bibinfo {volume} {88}},\ \bibinfo {pages}
  {035005} (\bibinfo {year} {2016})}\BibitemShut {NoStop}%
\bibitem [{\citenamefont {Sushkov}\ and\ \citenamefont
  {Castro~Neto}(2013)}]{PhysRevLett.110.186601}%
  \BibitemOpen
  \bibfield  {author} {\bibinfo {author} {\bibfnamefont {O.~P.}\ \bibnamefont
  {Sushkov}}\ and\ \bibinfo {author} {\bibfnamefont {A.~H.}\ \bibnamefont
  {Castro~Neto}},\ }\href {\doibase 10.1103/PhysRevLett.110.186601} {\bibfield
  {journal} {\bibinfo  {journal} {Phys. Rev. Lett.}\ }\textbf {\bibinfo
  {volume} {110}},\ \bibinfo {pages} {186601} (\bibinfo {year}
  {2013})}\BibitemShut {NoStop}%
\bibitem [{\citenamefont {Wang}\ \emph {et~al.}(2017)\citenamefont {Wang},
  \citenamefont {Scarabelli}, \citenamefont {Du}, \citenamefont {Kuznetsova},
  \citenamefont {Pfeiffer}, \citenamefont {West}, \citenamefont {Gardner},
  \citenamefont {Manfra}, \citenamefont {Pellegrini}, \citenamefont {Wind},\
  and\ \citenamefont {Pinczuk}}]{Wang:2017aa}%
  \BibitemOpen
  \bibfield  {author} {\bibinfo {author} {\bibfnamefont {S.}~\bibnamefont
  {Wang}}, \bibinfo {author} {\bibfnamefont {D.}~\bibnamefont {Scarabelli}},
  \bibinfo {author} {\bibfnamefont {L.}~\bibnamefont {Du}}, \bibinfo {author}
  {\bibfnamefont {Y.~Y.}\ \bibnamefont {Kuznetsova}}, \bibinfo {author}
  {\bibfnamefont {L.~N.}\ \bibnamefont {Pfeiffer}}, \bibinfo {author}
  {\bibfnamefont {K.~W.}\ \bibnamefont {West}}, \bibinfo {author}
  {\bibfnamefont {G.~C.}\ \bibnamefont {Gardner}}, \bibinfo {author}
  {\bibfnamefont {M.~J.}\ \bibnamefont {Manfra}}, \bibinfo {author}
  {\bibfnamefont {V.}~\bibnamefont {Pellegrini}}, \bibinfo {author}
  {\bibfnamefont {S.~J.}\ \bibnamefont {Wind}}, \ and\ \bibinfo {author}
  {\bibfnamefont {A.}~\bibnamefont {Pinczuk}},\ }\href {\doibase
  10.1038/s41565-017-0006-x} {\bibfield  {journal} {\bibinfo  {journal} {Nature
  Nanotechnology}\ } (\bibinfo {year} {2017}),\
  10.1038/s41565-017-0006-x}\BibitemShut {NoStop}%
\bibitem [{\citenamefont {Jin}\ \emph {et~al.}(2017)\citenamefont {Jin},
  \citenamefont {Jhi},\ and\ \citenamefont {Liu}}]{C7NR05325H}%
  \BibitemOpen
  \bibfield  {author} {\bibinfo {author} {\bibfnamefont {K.-H.}\ \bibnamefont
  {Jin}}, \bibinfo {author} {\bibfnamefont {S.-H.}\ \bibnamefont {Jhi}}, \ and\
  \bibinfo {author} {\bibfnamefont {F.}~\bibnamefont {Liu}},\ }\href {\doibase
  10.1039/C7NR05325H} {\bibfield  {journal} {\bibinfo  {journal} {Nanoscale}\
  }\textbf {\bibinfo {volume} {9}},\ \bibinfo {pages} {16638} (\bibinfo {year}
  {2017})}\BibitemShut {NoStop}%
\bibitem [{\citenamefont {Cao}\ \emph {et~al.}(2017)\citenamefont {Cao},
  \citenamefont {Zhao},\ and\ \citenamefont {Louie}}]{PhysRevLett.119.076401}%
  \BibitemOpen
  \bibfield  {author} {\bibinfo {author} {\bibfnamefont {T.}~\bibnamefont
  {Cao}}, \bibinfo {author} {\bibfnamefont {F.}~\bibnamefont {Zhao}}, \ and\
  \bibinfo {author} {\bibfnamefont {S.~G.}\ \bibnamefont {Louie}},\ }\href
  {\doibase 10.1103/PhysRevLett.119.076401} {\bibfield  {journal} {\bibinfo
  {journal} {Phys. Rev. Lett.}\ }\textbf {\bibinfo {volume} {119}},\ \bibinfo
  {pages} {076401} (\bibinfo {year} {2017})}\BibitemShut {NoStop}%
\bibitem [{\citenamefont {Castro~Neto}\ and\ \citenamefont
  {Guinea}(2007)}]{PhysRevB.75.045404}%
  \BibitemOpen
  \bibfield  {author} {\bibinfo {author} {\bibfnamefont {A.~H.}\ \bibnamefont
  {Castro~Neto}}\ and\ \bibinfo {author} {\bibfnamefont {F.}~\bibnamefont
  {Guinea}},\ }\href {\doibase 10.1103/PhysRevB.75.045404} {\bibfield
  {journal} {\bibinfo  {journal} {Phys. Rev. B}\ }\textbf {\bibinfo {volume}
  {75}},\ \bibinfo {pages} {045404} (\bibinfo {year} {2007})}\BibitemShut
  {NoStop}%
\bibitem [{\citenamefont {Liu}\ \emph {et~al.}(2009)\citenamefont {Liu},
  \citenamefont {Wang}, \citenamefont {Shi}, \citenamefont {Yang},\ and\
  \citenamefont {Liu}}]{PhysRevB.80.233405}%
  \BibitemOpen
  \bibfield  {author} {\bibinfo {author} {\bibfnamefont {W.}~\bibnamefont
  {Liu}}, \bibinfo {author} {\bibfnamefont {Z.~F.}\ \bibnamefont {Wang}},
  \bibinfo {author} {\bibfnamefont {Q.~W.}\ \bibnamefont {Shi}}, \bibinfo
  {author} {\bibfnamefont {J.}~\bibnamefont {Yang}}, \ and\ \bibinfo {author}
  {\bibfnamefont {F.}~\bibnamefont {Liu}},\ }\href {\doibase
  10.1103/PhysRevB.80.233405} {\bibfield  {journal} {\bibinfo  {journal} {Phys.
  Rev. B}\ }\textbf {\bibinfo {volume} {80}},\ \bibinfo {pages} {233405}
  (\bibinfo {year} {2009})}\BibitemShut {NoStop}%
\bibitem [{\citenamefont {Ouyang}\ \emph {et~al.}(2011)\citenamefont {Ouyang},
  \citenamefont {Peng}, \citenamefont {Liu},\ and\ \citenamefont
  {Liu}}]{doi:10.1021/nn200580w}%
  \BibitemOpen
  \bibfield  {author} {\bibinfo {author} {\bibfnamefont {F.}~\bibnamefont
  {Ouyang}}, \bibinfo {author} {\bibfnamefont {S.}~\bibnamefont {Peng}},
  \bibinfo {author} {\bibfnamefont {Z.}~\bibnamefont {Liu}}, \ and\ \bibinfo
  {author} {\bibfnamefont {Z.}~\bibnamefont {Liu}},\ }\href {\doibase
  10.1021/nn200580w} {\bibfield  {journal} {\bibinfo  {journal} {ACS Nano}\
  }\textbf {\bibinfo {volume} {5}},\ \bibinfo {pages} {4023} (\bibinfo {year}
  {2011})}\BibitemShut {NoStop}%
\bibitem [{\citenamefont {Lee}\ \emph {et~al.}(2013)\citenamefont {Lee},
  \citenamefont {Roy}, \citenamefont {Wohlwend}, \citenamefont {Varshney},
  \citenamefont {Ferguson}, \citenamefont {Mitchel},\ and\ \citenamefont
  {Farmer}}]{doi:10.1063/1.4807426}%
  \BibitemOpen
  \bibfield  {author} {\bibinfo {author} {\bibfnamefont {J.}~\bibnamefont
  {Lee}}, \bibinfo {author} {\bibfnamefont {A.~K.}\ \bibnamefont {Roy}},
  \bibinfo {author} {\bibfnamefont {J.~L.}\ \bibnamefont {Wohlwend}}, \bibinfo
  {author} {\bibfnamefont {V.}~\bibnamefont {Varshney}}, \bibinfo {author}
  {\bibfnamefont {J.~B.}\ \bibnamefont {Ferguson}}, \bibinfo {author}
  {\bibfnamefont {W.~C.}\ \bibnamefont {Mitchel}}, \ and\ \bibinfo {author}
  {\bibfnamefont {B.~L.}\ \bibnamefont {Farmer}},\ }\href {\doibase
  10.1063/1.4807426} {\bibfield  {journal} {\bibinfo  {journal} {Appl. Phys.
  Lett.}\ }\textbf {\bibinfo {volume} {102}},\ \bibinfo {pages} {203107}
  (\bibinfo {year} {2013})}\BibitemShut {NoStop}%
\bibitem [{\citenamefont {Po}\ \emph {et~al.}(2017)\citenamefont {Po},
  \citenamefont {Vishwanath},\ and\ \citenamefont {Watanabe}}]{Po:2017aa}%
  \BibitemOpen
  \bibfield  {author} {\bibinfo {author} {\bibfnamefont {H.~C.}\ \bibnamefont
  {Po}}, \bibinfo {author} {\bibfnamefont {A.}~\bibnamefont {Vishwanath}}, \
  and\ \bibinfo {author} {\bibfnamefont {H.}~\bibnamefont {Watanabe}},\ }\href
  {\doibase 10.1038/s41467-017-00133-2} {\bibfield  {journal} {\bibinfo
  {journal} {Nature Commun.}\ }\textbf {\bibinfo {volume} {8}},\ \bibinfo
  {pages} {50} (\bibinfo {year} {2017})}\BibitemShut {NoStop}%
\bibitem [{\citenamefont {Bradlyn}\ \emph {et~al.}(2017)\citenamefont
  {Bradlyn}, \citenamefont {Elcoro}, \citenamefont {Cano}, \citenamefont
  {Vergniory}, \citenamefont {Wang}, \citenamefont {Felser}, \citenamefont
  {Aroyo},\ and\ \citenamefont {Bernevig}}]{Bradlyn:2017aa}%
  \BibitemOpen
  \bibfield  {author} {\bibinfo {author} {\bibfnamefont {B.}~\bibnamefont
  {Bradlyn}}, \bibinfo {author} {\bibfnamefont {L.}~\bibnamefont {Elcoro}},
  \bibinfo {author} {\bibfnamefont {J.}~\bibnamefont {Cano}}, \bibinfo {author}
  {\bibfnamefont {M.~G.}\ \bibnamefont {Vergniory}}, \bibinfo {author}
  {\bibfnamefont {Z.}~\bibnamefont {Wang}}, \bibinfo {author} {\bibfnamefont
  {C.}~\bibnamefont {Felser}}, \bibinfo {author} {\bibfnamefont {M.~I.}\
  \bibnamefont {Aroyo}}, \ and\ \bibinfo {author} {\bibfnamefont {B.~A.}\
  \bibnamefont {Bernevig}},\ }\href {http://dx.doi.org/10.1038/nature23268}
  {\bibfield  {journal} {\bibinfo  {journal} {Nature}\ }\textbf {\bibinfo
  {volume} {547}},\ \bibinfo {pages} {298} (\bibinfo {year}
  {2017})}\BibitemShut {NoStop}%
\bibitem [{\citenamefont {Kruthoff}\ \emph {et~al.}()\citenamefont {Kruthoff},
  \citenamefont {de~Boer}, \citenamefont {van Wezel}, \citenamefont {Kane},\
  and\ \citenamefont {Slager}}]{combi}%
  \BibitemOpen
  \bibfield  {author} {\bibinfo {author} {\bibfnamefont {J.}~\bibnamefont
  {Kruthoff}}, \bibinfo {author} {\bibfnamefont {J.}~\bibnamefont {de~Boer}},
  \bibinfo {author} {\bibfnamefont {J.}~\bibnamefont {van Wezel}}, \bibinfo
  {author} {\bibfnamefont {C.~L.}\ \bibnamefont {Kane}}, \ and\ \bibinfo
  {author} {\bibfnamefont {R.-J.}\ \bibnamefont {Slager}},\ }\href@noop {}
  {\bibinfo  {journal} {arXiv:1612.02007}\ }\BibitemShut {NoStop}%
\bibitem [{\citenamefont {Noh}\ \emph {et~al.}()\citenamefont {Noh},
  \citenamefont {Benalcazar}, \citenamefont {Huang}, \citenamefont {Collins},
  \citenamefont {Chen}, \citenamefont {Hughes},\ and\ \citenamefont
  {Rechtsman}}]{C2}%
  \BibitemOpen
\bibfield  {journal} {  }\bibfield  {author} {\bibinfo {author} {\bibfnamefont
  {J.}~\bibnamefont {Noh}}, \bibinfo {author} {\bibfnamefont {W.~A.}\
  \bibnamefont {Benalcazar}}, \bibinfo {author} {\bibfnamefont
  {S.}~\bibnamefont {Huang}}, \bibinfo {author} {\bibfnamefont {M.~J.}\
  \bibnamefont {Collins}}, \bibinfo {author} {\bibfnamefont {K.}~\bibnamefont
  {Chen}}, \bibinfo {author} {\bibfnamefont {T.~L.}\ \bibnamefont {Hughes}}, \
  and\ \bibinfo {author} {\bibfnamefont {M.~C.}\ \bibnamefont {Rechtsman}},\
  }\href@noop {} {\bibinfo  {journal} {arXiv:1611.02373}\ }\BibitemShut
  {NoStop}%
\bibitem [{\citenamefont {Tang}\ \emph {et~al.}(2017)\citenamefont {Tang},
  \citenamefont {Zhang}, \citenamefont {Deng}, \citenamefont {Fan},
  \citenamefont {Zhang},\ and\ \citenamefont {Sun}}]{C6RA27465J}%
  \BibitemOpen
  \bibfield  {author} {\bibinfo {author} {\bibfnamefont {G.~P.}\ \bibnamefont
  {Tang}}, \bibinfo {author} {\bibfnamefont {Z.~H.}\ \bibnamefont {Zhang}},
  \bibinfo {author} {\bibfnamefont {X.~Q.}\ \bibnamefont {Deng}}, \bibinfo
  {author} {\bibfnamefont {Z.~Q.}\ \bibnamefont {Fan}}, \bibinfo {author}
  {\bibfnamefont {H.}~\bibnamefont {Zhang}}, \ and\ \bibinfo {author}
  {\bibfnamefont {L.}~\bibnamefont {Sun}},\ }\href {\doibase
  10.1039/C6RA27465J} {\bibfield  {journal} {\bibinfo  {journal} {RSC Adv.}\
  }\textbf {\bibinfo {volume} {7}},\ \bibinfo {pages} {8927} (\bibinfo {year}
  {2017})}\BibitemShut {NoStop}%
\bibitem [{\citenamefont {Yu}\ \emph {et~al.}(2008)\citenamefont {Yu},
  \citenamefont {Lupton}, \citenamefont {Liu}, \citenamefont {Liu},\ and\
  \citenamefont {Liu}}]{Yu2008}%
  \BibitemOpen
  \bibfield  {author} {\bibinfo {author} {\bibfnamefont {D.}~\bibnamefont
  {Yu}}, \bibinfo {author} {\bibfnamefont {E.~M.}\ \bibnamefont {Lupton}},
  \bibinfo {author} {\bibfnamefont {M.}~\bibnamefont {Liu}}, \bibinfo {author}
  {\bibfnamefont {W.}~\bibnamefont {Liu}}, \ and\ \bibinfo {author}
  {\bibfnamefont {F.}~\bibnamefont {Liu}},\ }\href {\doibase
  10.1007/s12274-008-8007-6} {\bibfield  {journal} {\bibinfo  {journal} {Nano
  Research}\ }\textbf {\bibinfo {volume} {1}},\ \bibinfo {pages} {56} (\bibinfo
  {year} {2008})}\BibitemShut {NoStop}%
\bibitem [{\citenamefont {Kresse}\ and\ \citenamefont
  {Hafner}(1993)}]{PhysRevB.47.558}%
  \BibitemOpen
  \bibfield  {author} {\bibinfo {author} {\bibfnamefont {G.}~\bibnamefont
  {Kresse}}\ and\ \bibinfo {author} {\bibfnamefont {J.}~\bibnamefont
  {Hafner}},\ }\href {\doibase 10.1103/PhysRevB.47.558} {\bibfield  {journal}
  {\bibinfo  {journal} {Phys. Rev. B}\ }\textbf {\bibinfo {volume} {47}},\
  \bibinfo {pages} {558} (\bibinfo {year} {1993})}\BibitemShut {NoStop}%
\bibitem [{\citenamefont {Kresse}\ and\ \citenamefont
  {Hafner}(1994)}]{PhysRevB.49.14251}%
  \BibitemOpen
  \bibfield  {author} {\bibinfo {author} {\bibfnamefont {G.}~\bibnamefont
  {Kresse}}\ and\ \bibinfo {author} {\bibfnamefont {J.}~\bibnamefont
  {Hafner}},\ }\href {\doibase 10.1103/PhysRevB.49.14251} {\bibfield  {journal}
  {\bibinfo  {journal} {Phys. Rev. B}\ }\textbf {\bibinfo {volume} {49}},\
  \bibinfo {pages} {14251} (\bibinfo {year} {1994})}\BibitemShut {NoStop}%
\bibitem [{\citenamefont {Kresse}\ and\ \citenamefont
  {Furthm{\"u}ller}(1996)}]{KRESSE199615}%
  \BibitemOpen
  \bibfield  {author} {\bibinfo {author} {\bibfnamefont {G.}~\bibnamefont
  {Kresse}}\ and\ \bibinfo {author} {\bibfnamefont {J.}~\bibnamefont
  {Furthm{\"u}ller}},\ }\href {\doibase
  http://dx.doi.org/10.1016/0927-0256(96)00008-0} {\bibfield  {journal}
  {\bibinfo  {journal} {Computational Materials Science}\ }\textbf {\bibinfo
  {volume} {6}},\ \bibinfo {pages} {15 } (\bibinfo {year} {1996})}\BibitemShut
  {NoStop}%
\bibitem [{\citenamefont {Kresse}\ and\ \citenamefont
  {Furthm\"uller}(1996)}]{PhysRevB.54.11169}%
  \BibitemOpen
  \bibfield  {author} {\bibinfo {author} {\bibfnamefont {G.}~\bibnamefont
  {Kresse}}\ and\ \bibinfo {author} {\bibfnamefont {J.}~\bibnamefont
  {Furthm\"uller}},\ }\href {\doibase 10.1103/PhysRevB.54.11169} {\bibfield
  {journal} {\bibinfo  {journal} {Phys. Rev. B}\ }\textbf {\bibinfo {volume}
  {54}},\ \bibinfo {pages} {11169} (\bibinfo {year} {1996})}\BibitemShut
  {NoStop}%
\bibitem [{\citenamefont {Perdew}\ \emph {et~al.}(1996)\citenamefont {Perdew},
  \citenamefont {Burke},\ and\ \citenamefont
  {Ernzerhof}}]{PhysRevLett.77.3865}%
  \BibitemOpen
  \bibfield  {author} {\bibinfo {author} {\bibfnamefont {J.~P.}\ \bibnamefont
  {Perdew}}, \bibinfo {author} {\bibfnamefont {K.}~\bibnamefont {Burke}}, \
  and\ \bibinfo {author} {\bibfnamefont {M.}~\bibnamefont {Ernzerhof}},\ }\href
  {\doibase 10.1103/PhysRevLett.77.3865} {\bibfield  {journal} {\bibinfo
  {journal} {Phys. Rev. Lett.}\ }\textbf {\bibinfo {volume} {77}},\ \bibinfo
  {pages} {3865} (\bibinfo {year} {1996})}\BibitemShut {NoStop}%
\end{thebibliography}
\end{document}